\documentclass[a4paper,10pt]{article}
\pdfoutput=1 
\usepackage[utf8]{inputenc}

\usepackage{amsmath, amssymb, amsthm}

\usepackage{graphicx, verbatim}

\usepackage[square, colon, numbers, sort&compress]{natbib}

\usepackage{chemarr}

\usepackage{authblk}

\newcommand{\der}{{\rm d}}
\newcommand{\ve}{\varepsilon}

\title{Gene expression noise is affected differentially by
 feedback in burst frequency and burst size}
\author[1]{Pavol Bokes}
\author[2]{Abhyudai Singh}
\affil[1]{Department of Applied Mathematics and Statistics, Comenius University, Bratislava, Slovakia}
\affil[2]{Department of Electrical and Computer Engineering, University of Delaware, Newark, Delaware, USA}

\date{}

\begin{document}

\maketitle

\begin{abstract}
Inside individual cells, expression of genes is stochastic across organisms ranging from bacterial to human cells. 
A ubiquitous feature of stochastic expression is burst-like synthesis of gene products, which drives considerable 
intercellular variability in protein levels across an isogenic cell population. One common mechanism by which cells 
control such stochasticity is negative feedback regulation, where a protein inhibits its own synthesis. For a single 
gene that is expressed in bursts, negative feedback can affect the burst frequency or the burst size. In order to 
compare these feedback types, we study a piecewise deterministic model for gene expression of a self-regulating gene. 
Mathematically tractable steady-state protein distributions are derived and used to compare the noise suppression 
abilities of the two feedbacks. Results show that in the low noise regime, both feedbacks are similar in term of 
their noise buffering abilities.  Intriguingly, feedback in burst size outperforms the feedback in burst frequency 
in the high noise regime. Finally, we discuss various regulatory strategies by which cells implement feedback to 
control burst sizes of expressed proteins at the level of single cells.
\end{abstract}

\section{Introduction}
\label{sec:intro}

Stochastic expression of genes drives significant random fluctuations (noise) in protein 
copy numbers over time in single 
cells~\citep{elowitz2002sge,bpm06,rao05,tcl10,bkc03,keb05,Eldar:2010kk,mno12}. These 
fluctuations manifest as  cell-to-cell
variability in level of a protein, even in genetically-identical populations under 
the same external 
conditions. Stochastic gene expression poses a challenge for the precise control 
of cellular function, placing cells under evolutionary pressure to minimize
the noise in vital proteins~\citep{lehner2008selection, singh2009evolution,lps07}. 
Not surprisingly, cell use diverse regulatory mechanisms to buffer noise in gene 
expression \citep{lvp10,sbf15,sak06,pep08,bhj03}. Negative feedback, by which the 
synthesis of gene products 
is  switched off in their excess, and switched on in their absence, is a commonly 
used mechanism for noise 
control~\citep{becskei2000engineering,swain2004efficient,kee01,sih09b,nam09,vob12,bandiera2016experimental}.

A major contributor to the overall noise in gene expression is the synthesis
of proteins in random bursts, and these bursts can occur both at the transcriptional and translation levels.
At the transcriptional level, a promoter can slowly become active, producing a burst of mRNAs before becoming 
inactive ~\citep{raj2006stochastic,suter2011mammalian,Bothma03072014,dar2012transcriptional,singh_transient_2014,ksk15}. 
At the translational level, a short-lived unstable mRNA degrades after synthesizing a burst of protein
molecules ~\citep{cai2006spe, yu2006probing, pau05}.  In the context of such burst-like gene expression, 
negative feedback can act either by reducing the frequency with which bursts occur, or by reducing
their size. 

Transcriptional control can reduce the frequency or the size of transcriptional bursts, the former by 
hindering promoter activation and the latter by enhancing promoter inactivation. By controlling 
transcription, the frequency of translational bursts can also be regulated; however, their size 
needs be regulated post-transcriptionally. For example, many RNA binding proteins reduce the size 
of translational bursts by shortening the half-life of 
their own mRNA \citep{kulo05,jbp14,kbg13,bub11,mpp13,rhs09}. As a specific example, 
splicing factors typically bind to their own pre-mRNA to create an alternatively 
spliced transcript that is degraded via non-sense mediated degradation \citep{bub11,rhs09}.

In this paper, we present a theoretical comparison of the feedback in burst frequency
and burst size with regards to their performance in protein noise reduction. We use 
a piecewise deterministic mathematical framework according to which any 
decrease (due to decay) in protein concentration is deterministic and continuous, 
and any increase (due to synthesis) occurs in randomly timed discontinuous jumps of 
random size~\citep{friedman2006lsd,mackey2011molecular,mackey2013dynamic,bokes2013transcriptional,
mackey2015limiting,lin2016bursting, lin2016gene,bokes2015protein, ochab2010bimodal, ochab2015transcriptional}. 
This framework yields explicit formulae for protein probability 
density functions. We utilize these formulae by (i) calculating key noise characteristics by
numerical integration and (ii) perform qualitative analysis of noise reduction performance
by asymptotic evaluation of integrals. 

The outline of the paper is as follows. First, we introduce the chosen modelling
framework in Section~\ref{sec:frame}. This is used to study feedback in burst frequency
in Section~\ref{sec:freq} and burst size in Section~\ref{sec:size}. Then follows 
a more technical Section~\ref{sec:asymptotics}, in which strong-feedback asymptotics of 
protein mean and noise are developed. The results of Sections~\ref{sec:frame}--\ref{sec:asymptotics},
and their implications, are summarised in a non-technical language in Section~\ref{sec:results}.
Finally, Section~\ref{sec:discuss} is devoted to discussing our results, especially in the context of
previous theoretical comparisons between different types of negative 
feedback~\citep{swain2004efficient,singh2011negative,zeiser2009hybrid,singh2009reducing,singh2011genetic,bandiera2016experimental}.

\section{Modelling framework}
\label{sec:frame}

We study a random telegraph model for stochastic gene expression with feedback in general form, 
\begin{equation}
 \label{random_telegraph}
 {\rm Off}\xrightleftharpoons[k_{\rm off}(x)]{k_{\rm on}(x)}{\rm On} \xrightarrow{k_{\rm p}(x)} 
   {\rm X} \xrightarrow{k_{\rm d}(x)} \emptyset,
\end{equation}
according to which the gene transitions between an inactive Off state and an On state, 
from which the protein X is synthesised, and eventually degraded. 

The reaction rates $k_{\rm on}(x)$ of activation, $k_{\rm off}(x)$ of inactivation, $k_{\rm p}(x)$ of protein production, 
and $k_{\rm d}(x)$ of degradation depend on the current amount $x$ of protein X in the system. We shall treat 
$x$ as a continuous quantity, i.e. a concentration, which evolves according to 
the ODE ${\der x}/{\der t}=k_{\rm p}(x)$ if the gene is On and according to 
${\der x}/{\der t}= -k_{\rm d}(x)$ if the gene is Off.

We shall assume that the inactivation rate $k_{\rm off}(x)$ and the protein synthesis rate $k_{\rm p}(x)$ are much
faster than the activation rate $k_{\rm on}(x)$ and decay rate $k_{\rm d}(x)$. In that case, the gene is mostly 
Off, while the protein level slowly decays, switching momentarily into the On state, upon which a short 
spell of rapid production of protein, i.e. a burst, ensues, during which the effect of degradation is negligible. Bursts can be either 
transcriptional, in which case On and Off represent the active and inactive promoter states~\citep[cf.][]{raj2006stochastic}, or translational, 
in which case On and Off are meant to indicate the presence or absence of an unstable mRNA transcript~\citep[cf.][]{lin2016gene}

In order to characterise the dynamics of a single burst, we denote by $y$ the protein concentration on entering 
the On state, and let $G(x,y)$, where $x>y$, be the probability that the protein concentration exceeds $x$ before the burst is terminated.

For any concentration level $z$ such that $x>z>y$, the ratio $\der z/ k_{\rm p}(z)$ gives the time of gene activity required to 
produce $\der z$ of protein, while $k_{\rm off}(z)$ gives the hazard rate for aborting the burst. The probability 
that it is not aborted before $x$ is reached is then determined by exponentiating the cumulative hazard rate~\citep[cf.][]{crudu2009hybrid, crudu2012convergence},
\begin{equation}
\label{kernel}
 G(x,y) = {\rm e}^{-\int_y^x \frac{k_{\rm off}(z)}{k_{\rm p}(z)} \der z }.
\end{equation}
If $k_{\rm p}(x)$ and $k_{\rm off}(x)$ are constants, then~\eqref{kernel} implies exponential distribution of burst sizes~\citep[cf.][]{friedman2006lsd}.
We assume that $\int^\infty  \frac{k_{\rm off}(z)}{k_{\rm p}(z)} \der z = \infty$ so that bursts are finite with probability one.

The probability density $p(x,t)$ of having $x$ protein at a time $t$ satisfies a continuity equation
\begin{equation}
\label{continuity}
 \frac{\partial p}{\partial t} + \frac{\partial J}{\partial x} = 0,
\end{equation}
where 
\begin{equation}
\label{flux}
 J = - k_d(x) p(x,t) 
  + \int_0^x G(x,y) k_{\rm on}(y) p(y, t)\der y.
\end{equation}
The term $J$ is the probability flux, which specifies how much probability mass passes through a given point $x$ 
(in the positive direction) per unit time. Equation~\eqref{continuity} mathematically expresses the fact that all changes in 
probability density function are due to this flux, i.e. that the total mass remains conserved. By~\eqref{flux}, the flux 
consists of a local term $-k_{\rm d}(x)p(x,t)$, which gives the transfer of probability mass due to protein decay; 
since decay leads to movement of probability mass in the negative direction, this term takes a negative sign. The other term
in~\eqref{flux} is nonlocal, and gives the transfer of probability mass due to bursts that start at a protein 
concentration $y$, which is lower than $x$, and end at a concentration which exceeds $x$. The probability of
a burst being of a sufficient size is equal to $G(x,y)$, which needs to be multiplied by the probability $k_{\rm on}(y) p(y,t)$ 
of the burst actually having been initiated, and then integrated over all possible starting 
concentrations $y$; this indeed yields the second term in~\eqref{flux}. Equations~\eqref{continuity}--\eqref{flux} are
equivalent to the Chapman--Kolmogorov differential equation~\citep{schuss2009theory} for a drift-jump (i.e. diffusion-less) Markov process, 
where the drift is due to protein degradation and jumps are due to bursts.

%this term takes a negative sign, since decay causes a movement of the probability mass 
%The total flux $J$ consists of the advective term $-k_{\rm d}(x)p(x,t)$, which is due to protein degradation
%and causes the probability mass to move in the direction of decreasing $x$, and a non-local term due to
%production in bursts. The non=

\begin{comment}
For a steady state solution $p(x,t)=p(x)$ to~\eqref{main-continuity}-\eqref{main-flux}, the probability flux $J$ as
well as its derivative ${\der J}/{\der x}$ is equal to zero
\begin{align}
 k_d(x) p(x) &=  \int_0^x {\rm e}^{-\int_y^x \frac{k_{\rm off}(z)}{k_{\rm p}(z)}\der z }k_{\rm on}(y) p(y)\der y,
    \label{ss_eq_1} \\
 \frac{\der}{\der x}(k_{\rm d}(x) p(x)) &=   k_{\rm on}(x) p(x) \nonumber\\
 - \frac{k_{\rm off}(x)}{k_{\rm p}(x)}& \int_0^x 
{\rm e}^{- \int_y^x \frac{k_{\rm off}(z)}{k_{\rm p}(z)}\der z} k_{\rm on}(y)p(y)\der y. \label{ss_eq_2} 
\end{align}
Multiplying~\eqref{ss_eq_1} by $k_{\rm off}(x)/k_{\rm p}(x)$ and adding the result to~\eqref{ss_eq_2} yields
\begin{equation}
 \label{ode}
 \frac{\der}{\der x}(k_{\rm d}(x) p(x)) + \frac{k_{\rm off}(x) k_{\rm d}(x) p(x)}{k_{\rm p}(x)} =  
   k_{\rm on}(x) p(x).
\end{equation}
Integrating the linear first-order ODE yields the formula~\eqref{main-ss} for the steady state protein distribution.
\end{comment}

The steady-state probability distribution $p(x)$ is found by setting $J=0$ and ${\der J}/{\der x}=0$, i.e. 
\begin{align}
 k_d(x) p(x) &=  \int_0^x {\rm e}^{-\int_y^x \frac{k_{\rm off}(z)}{k_{\rm p}(z)}\der z }k_{\rm on}(y) p(y)\der y,
    \label{ss_eq_1} \\
 \frac{\der}{\der x}(k_{\rm d}(x) p(x)) &=   k_{\rm on}(x) p(x)  - \frac{k_{\rm off}(x)}{k_{\rm p}(x)} \int_0^x 
{\rm e}^{- \int_y^x \frac{k_{\rm off}(z)}{k_{\rm p}(z)}\der z} k_{\rm on}(y)p(y)\der y. \label{ss_eq_2} 
\end{align}
Eliminating the integral term from~\eqref{ss_eq_1}--\eqref{ss_eq_2}, one obtains a linear first-order ordinary
differential equation
\[
 \frac{\der}{\der x}(k_{\rm d}(x) p(x)) =  \left(  \frac{k_{\rm on}(x)}{k_{\rm d}(x)} - \frac{k_{\rm off}(x) }{k_{\rm p}(x)} \right) k_{\rm d}(x) p(x),
\]
which implies
\begin{equation}
\label{ss}
 p(x) = \frac{C}{k_{\rm d}(x)} 
    {\rm exp}\left(  \int \left(\frac{k_{\rm on}(x)}{k_{\rm d}(x)} - \frac{k_{\rm off}(x)}{k_p(x)}\right) \der x \right),
\end{equation}
where $C$ is the normalisation constant. In order to guarantee that solutions to the master 
equation~\eqref{continuity}--\eqref{flux} converge, as time increases to infinity, to~\eqref{ss}, one needs to impose, in addition to the integrability
condition for~\eqref{ss}, a number of additional constraints on the reaction rates to exclude certain degenerate 
types of behaviour such as extinction due to sublinear decay or infinite waiting for the next burst~\citep[see][]{mackey2013dynamic}.
An alternative derivation of~\eqref{ss}, which extends to non-bursting regimes also, can be found in~\citep{hufton2016intrinsic}. Additional 
methodology, such as finding mean first passage times, for problems of this kind can be found in~\citep{lin2016gene}.

\section{Feedback in burst frequency}
\label{sec:freq}

In this section we assume that the burst frequency $k_{\rm on}(x)$  decreases with 
increasing protein concentration $x$, the decay rate $k_{\rm d}(x)$ is proportional to the concentration 
of protein, and the mean burst size is a constant; specifically, we set
\begin{equation}
\label{normal}
 k_{\rm on}(x) = \frac{\ve^{-1}}{1 + (x/K)^H},\quad
 k_{\rm d}(x) = x,\quad
 \frac{k_{\rm p}(x)}{k_{\rm off}(x)} = \ve,
\end{equation}
where the dissociation constant $K$ and cooperativity coefficient $H$ parametrise
the decreasing Hill-type dependence of the burst frequency on the protein level. The 
Hill function~\eqref{normal} can be viewed as a quasi-steady-state
approximation of a finer-grained regulation mechanism, in which individual protein 
molecules cooperatively bind to multiple binding sites at the promoter of an inactive gene, 
whereby they prevent its transition to the active state. Stochastic models for negative 
autoregulation in the presence of bursting, which explicitly include binding of protein 
to promoter, have been studied in~\citet{gronlund2013transcription} and \citet{kumar2014exact}.

Both time and concentration scales are already nondimensionalised 
in~\eqref{normal}. Time is measured in the
units of mean protein lifetime: the decay rate is 
equal to the concentration of the protein. Concentration
is measured in the units of its mean in the absence
of self-repression ($K\rightarrow\infty$): the
unrepressed (maximal) burst frequency $\ve^{-1}$ is
the reciprocal of the mean burst size $\ve$. The parameter $\ve$ characterises the noisiness in the
autoregulatory system. Small $\ve$ implies frequent and
small bursts, large $\ve$ implies infrequent and large 
bursts.

Inserting~\eqref{normal} into~\eqref{ss}, we find that the steady-state
distribution assumes a Wentzel--Kramers--Brillouin (WKB) form~\citep{bressloff2014stochastic}
\begin{equation}
 \label{wkb}
 p(x) = \frac{C{\rm e}^{-\frac{\varPhi(x)}{\ve}}}{x},
\end{equation}
where
\begin{equation}
 \label{phi1}
 \varPhi(x) = - \int \frac{\der x}{x (1 + (x/K)^H)} + x = \frac{{\rm ln}\left( 1 + (x/K)^H\right)}{H} - {\rm ln}x + x.
\end{equation}
The integration constant $C$, mean concentration $\langle x \rangle$
and the variance $\sigma^2$ can be computed by numerical integration of
\begin{equation}
\label{integrals}
 C = \left(\int_0^\infty\frac{{\rm e}^{-\frac{\varPhi(x)}{\ve}}}{x}\der x \right)^{-1} ,\quad
 \langle x \rangle = \int_0^\infty x p(x) \der x,\quad
 \sigma^2 = \int_0^\infty (x- \langle x \rangle)^2 p(x) \der x.
\end{equation}
Some care has to be taken when evaluating in the $\ve \ll 1$
regime the first integral of~\eqref{integrals}, which, due to 
the exponentially small term, can easily become 
smaller than any absolute error tolerance. Such problems can be 
circumvented e.g. by multiplying the integrand by a sufficiently large 
constant, such as ${\rm e}^{\varPhi(x_{\rm s})/\ve}$, where $x_{\rm s}$
is defined as detailed below.

A scale-free characteristic of protein noise is the coefficient
of variation defined by
\begin{equation}
 \label{cv2}
 {\rm CV}^2 = \frac{\sigma^2}{\langle x \rangle^2}.
\end{equation}
We shall compare the coefficient of variation of the regulated protein to 
that of a constitutively expressed protein with the same mean and burst size; 
this requires the burst frequency set to $\langle x \rangle/\ve$. In the
constitutive case, the protein concentration has a gamma distribution
with the shape parameter being equal to the burst frequency~\citep{friedman2006lsd}. The 
squared coefficient of variation of the gamma distribution is the reciprocal of 
its shape parameter and hence of the burst frequency. Thus, we define 
\begin{equation}
 \label{cv2_rel}
 {\rm CV}^2_{\rm rel} = \frac{\langle x \rangle }{\ve }{\rm CV}^2
\end{equation}
as the relative coefficient of variation.

In the small-noise regime ($\ve\ll1$), explicit asymptotic 
expression for the noise characteristics can be derived using
the linear noise approximation (LNA). Here we can obtain
the LNA results easily by expanding the integrals in~\eqref{integrals} 
using Laplace's method~\citep{nayfeh2008perturbation}.

\begin{figure}
 \begin{center}
  \includegraphics[width=6.0cm]{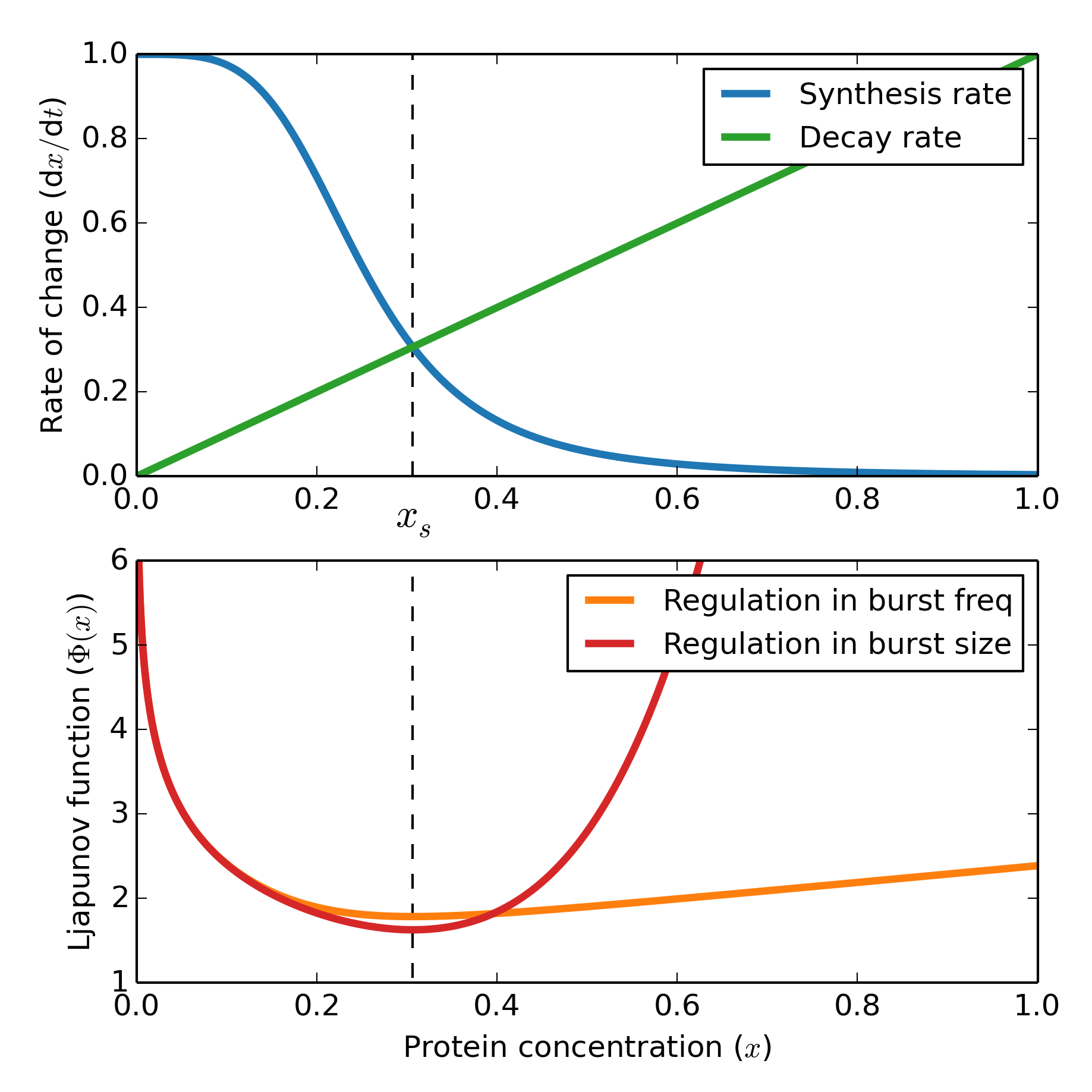}
 \end{center}
 \caption{Deterministic model and its two Lyapunov functions.
 The top panel shows the synthesis rate $(1+ (x/K)^H)^{-1}$ 
 and the decay rate $x$ as functions of $x$. The point $x_{\rm s}$ 
 at which they are equal is the steady state of the deterministic model
 (horizontal dashed lines). At the same point the Lyapunov functions
 are minimal (bottom panel). Note the flatness and asymmetry of the Lyapunov function
 used in the model for regulation via burst frequency in contrast with
 that used for regulation via burst size. The parameters are $H=4$, $K=1/4$.}
 \label{fig:potentials}
\end{figure}

The function $\varPhi(x)$ in~\eqref{wkb} is a Lyapunov function~\citep{strogatz2014nonlinear} corresponding to
the deterministic model
\begin{equation}
\label{deterministic}
 \frac{\der x}{\der t} = \frac{1}{1 + (x/K)^H} - x,
\end{equation}
i.e. $\varPhi(x)$ is minimal when $x=x_{\rm s}$, where
$x_{\rm s}$ is the single stable steady state 
of~\eqref{deterministic}, decreasing for $x<x_{\rm s}$
and increasing for $x>x_{\rm s}$ (cf. Fig~\ref{fig:potentials}).

For $\ve\ll 1$, the dominant contribution of probability mass in the 
probability density
function~\eqref{wkb} comes from the neighbourhood of $x_{\rm s}$,
around which we have
\begin{equation}
\label{taylor}
 \varPhi(x) = \varPhi(x_{\rm s}) + \frac{1}{2}\varPhi''(x_{\rm s})(x- x_{\rm s})^2 + \ldots
\end{equation}
Substituting the parabolic approximation~\eqref{taylor} into~\eqref{wkb}
and neglecting higher order terms in the usual manner~\citep{nayfeh2008perturbation}, we find that
the protein concentration is approximately normally distributed with
the moments given by 
\begin{equation}
 \label{lna1}
 \langle x \rangle \sim x_{\rm s},\quad \sigma^2 \sim \frac{\ve}{\varPhi''(x_{\rm s})}\quad \text{if }\ve\ll1.
\end{equation}
Notably, the steady-state distribution reduces for $\ve\rightarrow0$ to a point mass situated at the steady 
state $x_{\rm s}$ of the ODE model; in Appendix A we provide a more general argument that 
the (time-dependent) master equation itself reduces, as $\varepsilon$ tends to zero, to the ODE 
model~\eqref{deterministic}. 

In order to express the variance~\eqref{lna1} in terms of the model parameters, we 
differentiate the Lyapunov function~\eqref{phi1} twice, finding
\begin{equation}
 \label{2nd_derivative}
 \varPhi''(x_{\rm s}) 
      = \frac{H(1 - x_{\rm s}) + 1}{x_{\rm s}}.
\end{equation}
Using~\eqref{lna1} and~\eqref{2nd_derivative} in~\eqref{cv2} and~\eqref{cv2_rel}
we find that asymptotic approximations
\begin{equation}
 \label{cv2_lna}
 {\rm CV}^2 \sim \frac{\ve}{x_{\rm s}(H(1-x_{\rm s}) + 1)},\quad 
 {\rm CV}^2_{\rm rel} \sim \frac{1}{H(1-x_{\rm s}) + 1}
\end{equation}
hold for the coefficients of variation in the small-noise regime ($\ve\ll 1$).

\section{Feedback in burst size}
\label{sec:size}

In this section we focus on the case of
\begin{equation}
\label{abnormal}
 k_{\rm on}(x) = \ve^{-1},\quad
 k_{\rm d}(x) = x,\quad
 \frac{k_{\rm p}(x)}{k_{\rm off}(x)} = \frac{\ve}{1 + (x/K)^H}.
\end{equation}
In contrast with~\eqref{normal}, the burst frequency in~\eqref{abnormal}
is constant, but the burst size is regulated: the burst growth 
rate $k_{\rm p}/k_{\rm off}$ decreases with increasing protein concentration.
The functional form of the decrease is again that of a Hill function parametrised
by $H$ and $K$. Small $\ve$ corresponds to small-noise regime.

If we interpret the Off/On states as indicators of translational, rather than promoter, 
activity, meaning that Off refers to the absence of transcripts and On indicates
the presence of a short-lived mRNA copy~\citep[cf.][]{lin2016gene}, then 
$k_{\rm off}(x)$ acquires the meaning of the mRNA degradation rate. The Hill-type
dependency in~\eqref{abnormal} can be achieved by an RNA-binding protein, 
which cooperatively catalyses the removal of its mRNA~\citep{kulo05,jbp14,kbg13}.

Inserting~\eqref{abnormal} into~\eqref{ss}, we find that the 
WKB form~\eqref{wkb} is still valid for the steady-state distribution, but
with a different Lyapunov function
\begin{equation}
\label{phi2}
 \varPhi(x) = \frac{x^{H+1}}{(H+1)K^H} - {\rm ln}x + x.
\end{equation}
The difference in the two Lyapunov functions reflects the difference
in the two stochastic models. However, both Lyapunov functions correspond to 
the same ordinary differential equation~\eqref{deterministic}, being
minimal at the ODE's steady state $x_{\rm s}$ (cf. Fig~\ref{fig:potentials}).
Thus, regardless of whether the feedback acts on
burst size or burst frequency, the steady-state protein concentration is 
narrowly distributed around the deterministic steady state $x_{\rm s}$ in the 
small-noise regime. In Appendix A, we provide a more general result 
which holds also outside of the steady-state regime: we show that
the master equation~\eqref{continuity}--\eqref{flux} reduces
for $\ve\rightarrow0$ to the ODE model~\eqref{deterministic},
regardless of whether the feedback acts via burst frequency~\eqref{normal} or 
burst size~\eqref{abnormal}.

Formulae~\eqref{wkb},~\eqref{integrals} and~\eqref{cv2} can be reused with the
new definition~\eqref{phi2} of $\varPhi$ to compute numerically the mean, 
variance, and the squared coefficient of variation of the protein distribution. 
However, a modification is due in the definition of the relative coefficient of variation.
Since the burst frequency is constant but the burst size is regulated, we
compare the CV$^2$ of a regulated protein to that of a constitutively expressed protein
with the same mean $\langle x\rangle$ and burst frequency $\ve^{-1}$,
adjusting the burst size to $\ve/\langle x \rangle$ as required.

The reciprocal $\ve$ of the burst frequency gives the squared 
coefficient of variation for the referential constitutively expressed 
protein. Thus, we define
\begin{equation}
 \label{cv2_rel_2}
 {\rm CV}^2_{\rm rel} = \frac{{\rm CV}^2}{\ve}
\end{equation}
as the relative coefficient of variation. This differs from~\eqref{cv2_rel},
in which the CV$^2$ of a protein with a regulated burst frequency 
was compared to the CV$^2$ of a constitutively expressed protein with the same 
mean and burst size, adjusting the burst frequency as required.

In the small-noise regime ($\ve\ll1$), the mean and variance satisfy~\eqref{lna1}, in which
the second derivative of the Lyapunov function is given not by~\eqref{2nd_derivative} but
\begin{equation}
\label{2nd_derivative_2}
 \varPhi''(x_{\rm s}) 
    = \frac{H(1 - x_{\rm s}) + 1}{x_{\rm s}^2},
\end{equation}
as is easily checked by differentiating~\eqref{phi2} twice.
Using~\eqref{lna1} and~\eqref{2nd_derivative_2} in the definitions of the
CV$^2$~\eqref{cv2} and the relative CV$^2$~\eqref{cv2_rel_2}, we find that
\begin{equation}
\label{cv2_lna_2}
 {\rm CV}^2 \sim \frac{\ve}{H(1-x_{\rm s}) + 1},\quad 
 {\rm CV}^2_{\rm rel} \sim \frac{1}{H(1-x_{\rm s}) + 1}
\end{equation}
hold in the small-noise regime in the case of regulation via burst size.

\section{Strong feedback asymptotics}
\label{sec:asymptotics}

Here we present an additional asymptotic analysis that yields explicit
predictions for mean and CV$^2$ that hold even in the large-noise
regime ($\ve=O(1)$), provided that the feedback is very strong ($K\ll\ve$).
We focus solely on the case of feedback in burst frequency, for which the 
strong-feedback asymptotics are more interesting than for feedback in 
burst size. The latter is nevertheless treated in Appendix B.

By~\eqref{wkb}--\eqref{phi1}, we have
\begin{equation}
 p(x) = C {\rm e}^{-x/\ve} x^{\frac{1}{\ve}-1} \left(1 + (x/K)^H\right)^{-\frac{1}{\ve H}}
\end{equation}
for the protein pdf.

The protein moments are given by
\begin{equation}
 \langle x^n \rangle = \frac{B_n}{B_0},
\end{equation}
where
\begin{align}
 B_n &= \int_0^\infty {\rm e}^{-x/\ve} x^{\frac{1}{\ve}-1+ n} \left(1 + (x/K)^H\right)^{-\frac{1}{\ve H}} \der x\label{bn}.
\end{align}
Note that $B_0^{-1}=C$ is the normalisation constant. Substituting $x = Ky$ in~\eqref{bn} yields
\begin{equation}
\label{bn_and_an}
 B_n = K^{\frac{1}{\ve}+n} A_n,
\end{equation}
where
\begin{equation}
\label{an}
A_n = \int_0^\infty {\rm e}^{-K y/\ve} y^{\frac{1}{\ve}-1+n} (1 + y^H)^{-\frac{1}{\ve H}} \der y.
\end{equation}
The protein mean and the squared coefficient of variation can be expressed
in terms of $A_n$, $n=0,1,2$, as
\begin{equation}
 \label{meancv_from_an}
 \langle x \rangle = \frac{K A_1}{A_0},\quad {\rm CV}^2 = \frac{\langle x^2 \rangle}{\langle x \rangle^2} - 1 = \frac{B_0 B_2}{B_1^2} - 1 = \frac{A_0 A_2}{A_1^2} - 1.
\end{equation}
Thus, we need to establish the limiting behaviour of Laplace transforms
\begin{equation}
A_n = \int_0^\infty {\rm e}^{-\lambda y} f_n(y) \der y, \quad\text{where}\quad f_n(y)=y^{\frac{1}{\ve}-1+n} (1 + y^H)^{-\frac{1}{\ve H}}
\end{equation}
for small values of the Laplace variable $\lambda=K/\ve$.

If $n\geq1$, then $\lambda\ll1$ implies $y\gg1$, so that $f_n(y)\sim y^{-1+n}$, and
\begin{equation}
\label{an_asympt}
 A_n \sim \int_0^\infty {\rm e}^{-\lambda y} y^{-1+n} \der y = (n-1)! \lambda^{-n}.
\end{equation}
The case of $n=0$ is an exception because of the divergence of the exponential
integral. 

For $n=0$ we split the integration range~\citep[see][]{hinch1991perturbation}
\begin{equation}
\label{rozklad}
 A_0 = \int_0^\infty {\rm e}^{-\lambda y} f_0(y) \der y = \int_0^\delta {\rm e}^{-\lambda y} f_0(y) \der y
	+ \int_\delta^\infty {\rm e}^{-\lambda y} f_0(y) \der y,
\end{equation}
where $\delta$ is chosen so that
\begin{equation}
1\ll \delta \ll \frac{1}{\lambda}
\end{equation}
is asymptotically satisfied.

%i.e. $\delta$ is chosen from an intermediate asymptotic region.

In the second integral of~\eqref{rozklad}, $y>\delta\gg1$ implies $f_0(y)=y^{\ve^{-1}-1}(1+y^H)^{-1/\ve H}\sim y^{-1}$. i.e.
\begin{equation}
 \label{first_chunk}
 \int_\delta^\infty {\rm e}^{-\lambda y} f_0(y) \der y \sim
 \int_\delta^\infty \frac{{\rm e}^{-\lambda y}}{y} \der y 
 = E_1(\lambda\delta) \sim - {\rm ln}\delta - {\rm ln}\lambda - \gamma,
\end{equation}
where $E_1(z)$ is the exponential integral and the right-hand side of~\eqref{first_chunk} is made of the first two terms
of its asymptotic expansion: $\gamma=0.577\ldots$ is the Euler--Mascheroni constant~\citep{abramowitz1972hmf}.

In the first integral on the right-hand side of~\eqref{rozklad}, we have $\lambda y < \lambda \delta \ll 1$, so that
\begin{align}
  \int_0^\delta {\rm e}^{-\lambda y} f_0(y) \der y & \sim \int_0^\delta f_0(y) \der y\nonumber\\
      &= \int_0^\delta y^{\frac{1}{\ve}-1} (1 + y^H)^{-\frac{1}{\ve H}} \der y. \label{second_chunk}
\end{align}
Substitution $v = y^H/(1+y^H)$ in~\eqref{second_chunk} yields
\begin{equation}
 \label{substitution}    
 \int_0^\delta y^{\frac{1}{\ve}-1} (1 + y^H)^{-\frac{1}{\ve H}} \der y
      = \frac{1}{H}\int_0^{\delta^H/(1+\delta^H)} v^{\frac{1}{\ve H} - 1} (1 - v)^{-1} \der v.
\end{equation}
Next, we extricate the divergent logarithmic part from the right-hand side of~\eqref{substitution} and 
neglect small terms in the convergent remainder (bearing in mind that $\delta\gg1$),
\begin{align}
 &\frac{1}{H}\int_0^{\delta^H/(1+\delta^H)} v^{\frac{1}{\ve H} - 1} (1 - v)^{-1} \der v\nonumber\\
 &\quad= \frac{1}{H}\left( {\rm ln}(1 + \delta^H) - 
  \int_0^{\delta^H/(1+\delta^H)}  \frac{1 - v^{\frac{1}{\ve H} - 1}}{1 - v} \der v \right)\nonumber\\
 &\quad\sim {\rm ln}\delta - 
  \frac{1}{H}\int_0^1  \frac{1 - v^{\frac{1}{\ve H} - 1}}{1 - v} \der v.\label{euler}
\end{align}
The integral on the right-hand side of~\eqref{euler} is Euler's integral representation of the $(\frac{1}{\ve H}-1)$-th
harmonic number~\citep{sandifer2007euler}, for which we have
\begin{equation}
\label{digamma}
\int_0^1  \frac{1 - v^{\frac{1}{\ve H} - 1}}{1 - v} \der v = \gamma + \psi\left(\frac{1}{\ve H}\right),
\end{equation}
where $\gamma$ is the Euler--Mascheroni constant and $\psi(s)$ is the digamma function (the logarithmic
derivative of the gamma function)~\citep{abramowitz1972hmf}. Thus, equations~\eqref{second_chunk}--\eqref{digamma} imply that
\begin{equation}
\label{second_chunk_2}
 \int_0^\delta {\rm e}^{-\lambda y} f_0(y) \der y \sim  {\rm ln}\delta  - \frac{1}{H}\left(\gamma + \psi\left(\frac{1}{\ve H}\right)\right)
\end{equation}
holds for the first integral on the right-hand side of~\eqref{rozklad}.

Inserting the approximations~\eqref{first_chunk} and~\eqref{second_chunk_2} into~\eqref{rozklad}, we obtain
\begin{equation}
\label{a0_asympt}
 A_0 \sim -{\rm ln}\lambda - q,
\end{equation}
where the constant $q$ is given by
\begin{equation}
 \label{q}
q = \gamma\left(1 + \frac{1}{H}\right) + \frac{1}{H}\psi\left(\frac{1}{\ve H}\right).
\end{equation}
The constant $q$ in~\eqref{a0_asympt} asymptotically dominated by the divergent logarithmic term 
$-{\rm ln}\lambda$ as $\lambda$ tends to zero; nevertheless, from a practical 
viewpoint, the constant is not negligible since the slowly convergent logarithmic 
term is in most situations comparable in magnitude.
%~\footnote{Quoting~\citet{arnold1984odu}, ``One should differentiate between the concepts of mathematical and physical 
%convergence $\ldots$ In practice, one can often treat logarithms as constants in asymptotics.''}

Using the asymptotic expressions~\eqref{an_asympt} for $A_1$ and $A_2$ and~\eqref{a0_asympt}
for $A_0$, together with the definition $\lambda=K/\ve$, in the formulae for the mean and 
CV$^2$~\eqref{meancv_from_an}, we find
\begin{equation}
\label{mean_cv2_log}
 \langle x \rangle \sim \frac{\ve}{{\rm ln}\frac{\ve}{K} - q},\quad {\rm CV}^2 \sim {\rm ln}\frac{\ve}{K} - q - 1,
\end{equation}
where the constant $q$ is given by~\eqref{q}; these expressions are valid in the strong feedback regime ($K\ll\ve$).
Additionally, we have
\begin{equation}
{\rm CV}^2_{\rm rel}  = \frac{\langle x \rangle}{\ve} {\rm CV}^2 \sim 1 - \frac{1}{{\rm ln}\frac{\ve}{K} - q}
\end{equation}
for the ratio ${\rm CV}^2_{\rm rel}$ of the regulated protein's CV$^2$ and that of a constitutively expressed protein with an equal mean 
expression and mean burst size.

It is interesting to compare the ultimate asymptotics~\eqref{mean_cv2_log} in the strong feedback regime
to the intermediate asymptotics obtained by taking $K$ small in the LNA results. In the latter case, 
the protein mean is approximated by the deterministic steady state $x_{\rm s}$, which is equal to the 
fixed point of the function $(1+(x/K)^H)^{-1}$. One sees easily that
\begin{equation}
 \label{powerlaw}
 x_{\rm s} \sim K^{\frac{H}{1+H}},\quad K\ll1,
\end{equation}
which suggests a faster, power-law, decrease in the protein
mean and, if inserted in~\eqref{lna1}--\eqref{cv2_lna}, a power-law increase in the 
coefficient of variation. However, the power-law mode is applicable
only in the low noise scenario for an intermediate range of $K$; as $K$ further
decreases, the logarithmic law~\eqref{mean_cv2_log} applies.

\section{Results}
\label{sec:results}

The methods described in the previous sections are used here to study 
the protein distributions and noise characteristics as a
function of strengthening negative feedback, whereby we shall
distinguish and juxtapose two cases, the first being the regulation
of the burst frequency and the second the regulation of the burst 
size.

The feedback strength is determined by one key parameter, the dissociation constant $K$,
which is defined as the concentration of protein required to achieve $50\%$
repression. The lower the dissociation constant, the lower the
concentration threshold for effective self-control: the stronger 
the feedback.

The dissociation constant is measured in the units of protein
concentration. In this study, the chosen unit of concentration 
is equal to the mean protein abundance in the absence of 
regulation. This natural choice of scale helps minimise the 
dimension of the parameter space of our models.

In addition to the dimensionless dissociation constant $K$, 
two other key parameters are identified: the cooperativity
coefficient $H$ and the noise parameter $\ve$. The cooperativity 
coefficient determines the steepness of the 
regulatory response to increasing protein concentrations. All 
examples in this study use $H=4$, which we consider a 
satisfactory representative for any $H>1$. The noncooperative
case $H=1$ is an exception and is treated in Appendix C.
Negative cooperativity ($0<H<1$) is not considered.

The noise parameter $\ve$ determines the size of bursts of 
protein synthesis in the chosen units of protein concentration.
The burst frequency is therefore $O(\ve^{-1})$ in order that
protein concentration be $O(1)$ as stated. In the small-noise
regime ($\ve\ll1$), analytically tractable expressions for
protein noise are obtained using linear noise approximation.
These will be contrasted with exact (i.e. not asymptotic) numerical
results.

\begin{figure}
\begin{center}
\includegraphics[width=10cm]{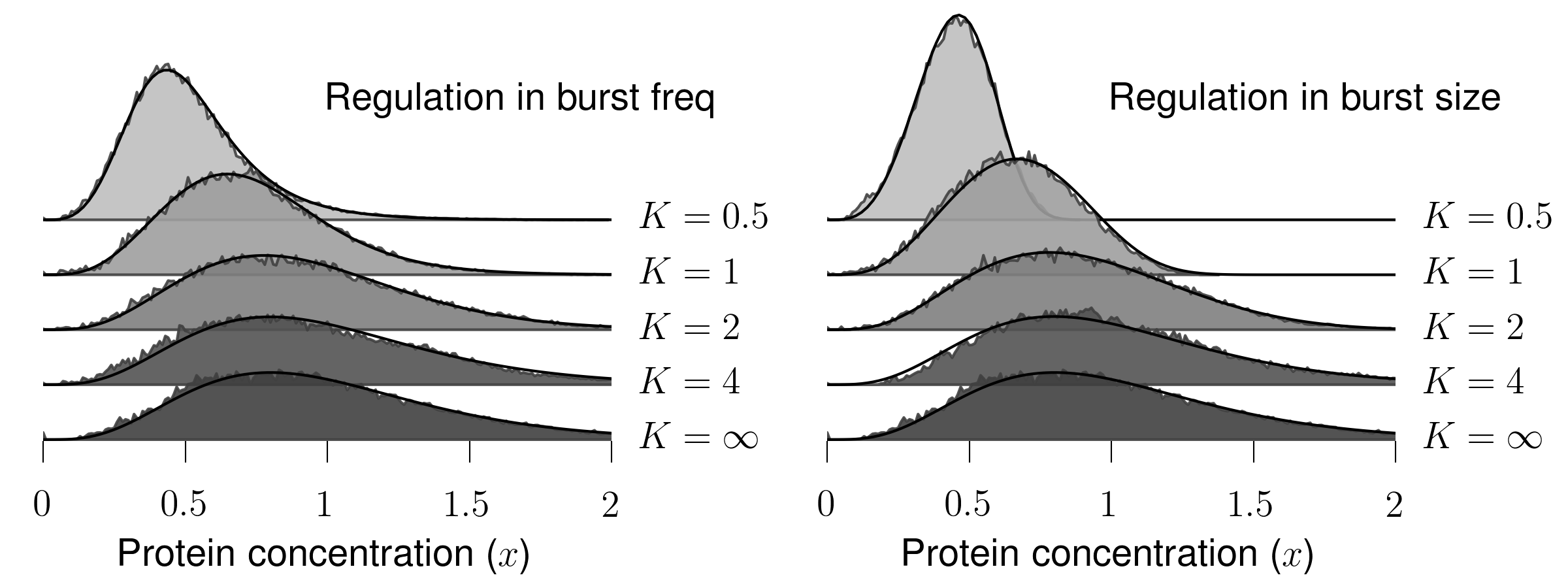}
\end{center}
\caption{Protein distributions for varied feedback strength. $\ve=0.2$.}
\label{fig:distributions}
\end{figure}

\paragraph{Right tails of protein distributions are narrower for feedback in burst size.}
The response of steady-state protein probability densities to increasing strength of 
either kind of feedback is investigated in Figure~\ref{fig:distributions}.
The exact result~\eqref{wkb}, in which the Lyapunov function $\varPhi$ is given
by~\eqref{phi1} for feedback in burst frequency and~\eqref{phi2} for feedback
in burst size, is shown in solid lines, and is compared to histograms
obtained by large-scale Gillespie simulations of a finer-grained discrete
stochastic model (description of which is found in Appendix D).

Inspecting the distributions in Figure~\ref{fig:distributions}, we infer that 
strengthening negative feedback (decreasing the dissociation constant $K$) 
of either kind reduces the mode and the width of steady state protein distributions. 
However, feedback in burst size of medium to high strength ($K=0.5$) leads to 
narrower distributions, in particular in their right tail, than feedback in burst 
frequency.  Such differences can intuitively be explained: negative regulation 
in burst  frequency leads to less frequent bursts, which are nevertheless large 
and contribute towards the right tail; regulation in burst size, on the other
hand, leads to smaller burst sizes, thus effectively reducing the tail.

\begin{figure}
\begin{center}
 \includegraphics[width=10cm]{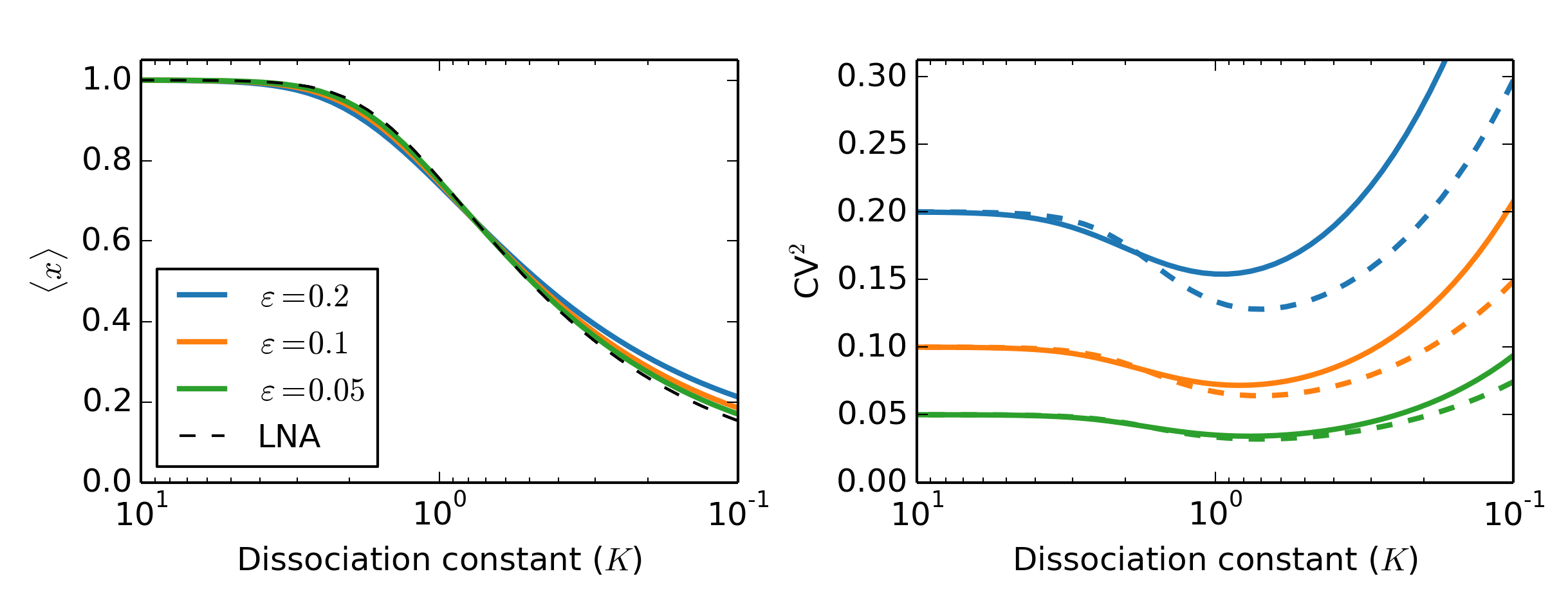}
\end{center}
\caption{Protein mean and CV$^2$ in response to strengthening feedback in burst frequency.}
\label{fig:cv2_bfreq_reg}
\end{figure}

\paragraph{Noise increases after an initial decrease in response to
 strengthening feedback in burst frequency.}
The impact of increasing the strength of feedback in burst frequency on 
protein mean and the squared coefficient of variation~\eqref{cv2} (CV$^2$)
is examined in Figure~\ref{fig:cv2_bfreq_reg}.  The horizontal axis in 
Figure~\ref{fig:cv2_bfreq_reg} gives the 
dissociation constant $K$ on the inverse logarithmic scale, i.e. moving constantly 
to the right along the axis corresponds to increasing feedback strength exponentially. 
The values of $K$ range from $K=10$ (low feedback strength) to $K=10^{-1}$ (strong feedback).

Solid lines give exact (as opposed to asymptotic) results obtained by numerical 
integration of the moments of the density~\eqref{wkb}--\eqref{cv2} for a selection of
values of $\ve$. Dashed lines give the linear noise approximation (LNA) results~\eqref{lna1}--\eqref{cv2_lna}, 
which are valid asymptotically in the small noise regime of $\ve\ll1$.

Focusing at first on the left panel of Figure~\ref{fig:cv2_bfreq_reg}, we observe that
the protein mean $\langle x\rangle$ monotonically decreases from $1$ 
(for $K=\infty$, i.e. without feedback) down to $0$ (for $K=0$, i.e. 
complete repression). In small- to moderate-noise regimes of $\ve$, the exact 
protein mean differs little from the LNA, which is equal to the steady 
state $x_{\rm s}$ of the deterministic model~\eqref{deterministic}.
The deterministic steady state is computed numerically as a unique fixed point 
of the production rate function $(1+(x/K)^H)^{-1}$.

Looking at the right panel of Figure~\ref{fig:cv2_bfreq_reg}, we see that
in the absence of regulation ($K=\infty$), we have CV$^2=\ve$. In response 
to lowering the dissociation constant $K$, the CV$^2$ first decreases and then 
increases back again. The LNA suggests that CV$^2$ goes to infinity as $K$ 
decreases. For $\ve\ll1$, the minimal CV$^2$ is achieved for $x_{\rm s}=(H+1)/2H$ 
and is equal to $4H\ve/(H+1)^2$. Comparing the LNA to the exact results,
we observe that larger values of $\ve$ make the initial
blip in the CV$^2$ less pronounced than the LNA predicts.

The initial decrease of noise in response to strengthening feedback strength
is intuitively expected: whenever protein is in surplus, additional synthesis
of protein is restricted by negative feedback, thus reducing deviations from
the mean. However,  as well as reducing deviations from the mean, the present 
type of feedback decreases the overall burst frequency, which implies higher
levels of noise.  The increase in noise due to lower burst frequencies demonstrably 
dominates over the decrease due to reduction in deviations from the mean for large 
feedback strengths. Below, we describe a more refined measure of noise, the relative
coefficient of variation, which adjusts for the decrease in overall burst frequency.
Additionally, at very low frequencies feedback in burst frequency may lose its ability 
to control protein fluctuations, whereby each burst overshoots and is followed by a 
period of complete self-repression, which takes a long time until enough protein is 
degraded, so that another burst may occur (which, inevitably, overshoots again).

\begin{figure}
\begin{center}
\includegraphics[width=10cm]{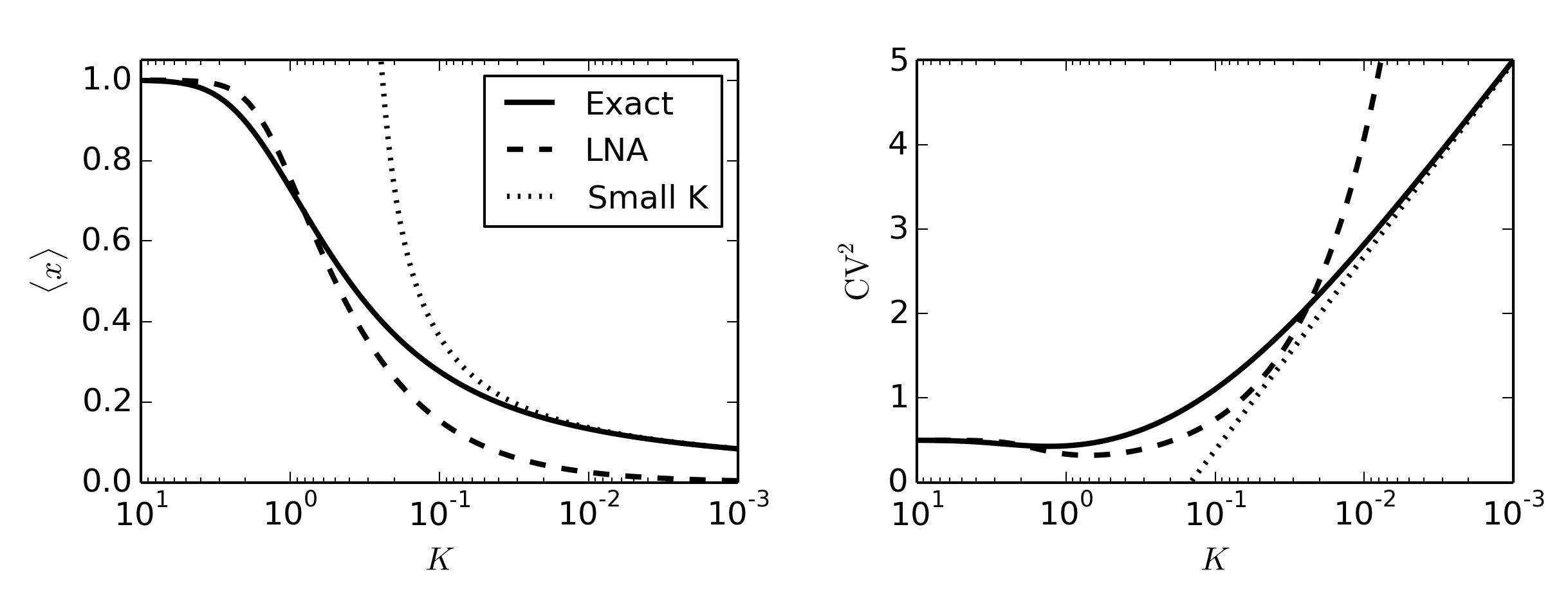}
\end{center}
\caption{Impact of strengthening feedback in burst frequency on mean and CV$^2$ of a noisy protein ($\ve=0.5$). 
Comparison of exact results (full line) with LNA (dashed line) and small $K$ (dotted line) asymptotics.}
\label{fig:asymptotics}
\end{figure}

\paragraph{LNA underestimates the mean and overestimates CV$^2$ of a noisy protein subject to strong 
feedback in burst frequency.} As we pointed out above, the LNA predicts that protein CV$^2$ 
diverges to infinity as the dissociation constant $K$ tends to zero. Additionally, it predicts
a power-law growth of the CV$^2$, which is due to a power-law decay of the mean~\eqref{powerlaw}. However, 
since increasing the feedback strength in burst frequency leads to large levels of noise, the 
LNA prediction, which assumed little noise, becomes ever less reliable as $K\to0$. Hence,
even for small $\ve$ the LNA will ultimately fail provided that the feedback is raised to
a sufficient strength.

For these reasons, we derived in Section~\ref{sec:asymptotics} asymptotic expressions for 
protein mean and CV$^2$ which are valid in the strong feedback regime even at high noise levels. 
More precisely, they are valid for $K\ll\ve$: noisier proteins require lesser feedback strengths 
for these results to apply. In contrast with the LNA prediction, a slow logarithmic 
decrease~\eqref{mean_cv2_log} in the mean and increase in the CV$^2$ is discovered. The LNA
power-law prediction can thus only be taken as an intermediate asymptotic result applicable
for small $\ve$ for intermediate ranges of feedback strengths.

The exact numerics, the linear-noise, and the strong-feedback asymptotics for protein mean and
CV$^2$ are compared in Figure~\ref{fig:asymptotics} for a relatively noisy protein $\ve=0.5$ (this 
value corresponds to a maximum of average two bursts per protein lifetime). 
Since $K$ is measured on a logarithmic scale, the limiting logarithmic dependence~\eqref{mean_cv2_log} of 
the CV$^2$ on $K$ (right-panel, dotted line) looks like a straight line, whose slope is
${\rm ln}10$ and intercept is ${\rm ln}\ve - q - 1$, where $q$ is defined by~\eqref{q}.

\begin{figure}
\begin{center}
 \includegraphics[width=10cm]{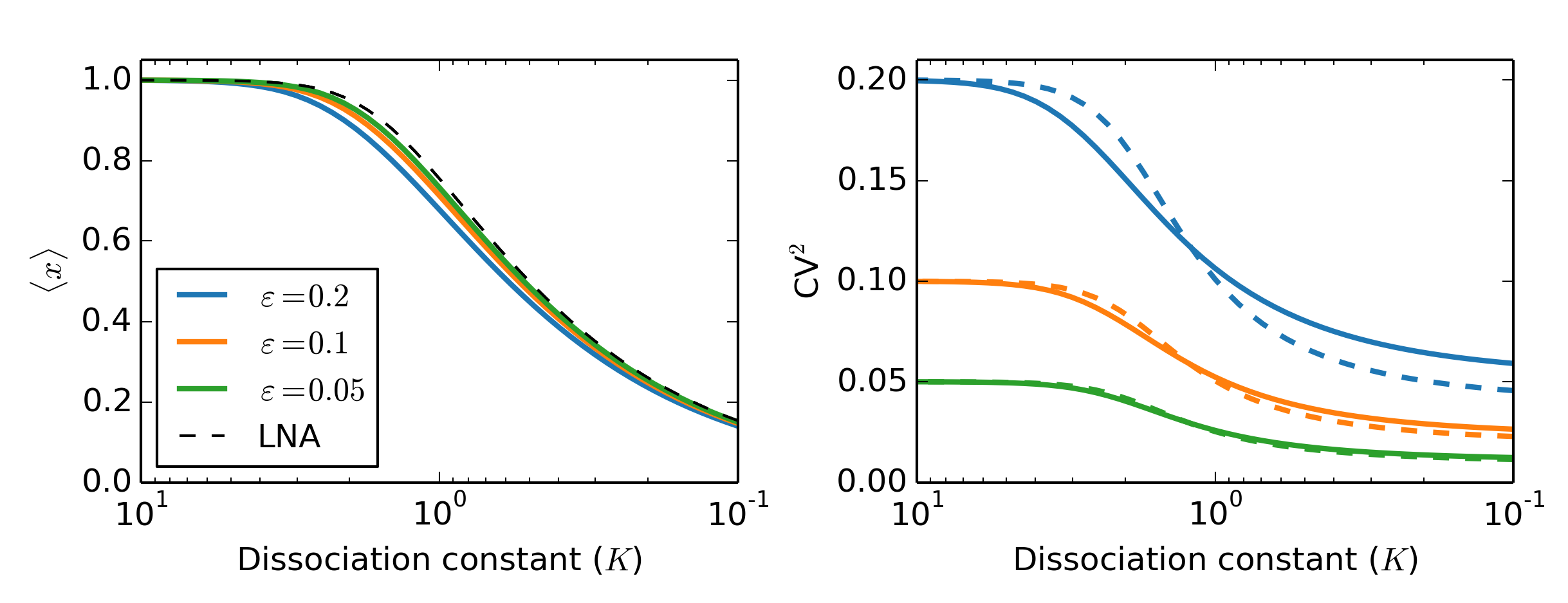}
\end{center}
\caption{Protein mean and CV$^2$ in response to strengthening feedback in burst size.}
\label{fig:cv2_bsize_reg}
\end{figure}
 
\paragraph{Noise decreases in response to strengthening feedback
in burst size.} In case of feedback in burst size, equations~\eqref{wkb},~\eqref{integrals}, and~\eqref{cv2}
are used  with the alternative definition~\eqref{phi2} of the Lyapunov function $\varPhi$ 
to compute the exact mean and CV$^2$ numerically. The LNA of the mean is again equal to 
the deterministic steady state $x_{\rm s}$, and the LNA of the CV$^2$ is given 
by~\eqref{cv2_lna_2}. 

In contrast to the previous case, the CV$^2$ decreases monotonically from
the value $\ve$ in the absence of regulation to a lower value in the limit of complete
repression, which is equal to $\ve/(H+1)$ in the small-noise regime. For moderate values of $\ve$,
the decrease in the CV$^2$ is less sigmoidal than predicted by the LNA.

\paragraph{Comparing noise of a regulated protein to that of a
constitutively expressed one with the same mean.} The ability
of negative feedback to suppress protein noise can be evaluated
by comparing the regulated protein CV$^2$ to that of a constitutively
expressed protein with the same mean level of expression. We refer 
to the ratio of the regulated CV$^2$ and constitutive CV$^2$
as the relative CV$^2$, or shortly ${\rm CV}^2_{\rm rel}$.

In our modelling framework, a constitutive expression of a protein 
is modulated by two parameters: the average burst size, which is measured 
in the chosen units of concentration; and the burst frequency, which
is measured in the units of protein decay rate constant. Our condition
of equal means implies that the product of these two must be equal 
to the mean of the regulated protein.

Having made requirement of equal means, one degree of freedom still
remains in the parameter space of the constitutively expressed protein,
and with this breadth of freedom a continuum of values of CV$^2$ can be attained. Therefore,
an extra condition is required on the constitutive protein to arrive at
a well-defined comparison.

This extra condition differs depending whether we investigate 
feedback in burst frequency or feedback in burst size. If feedback
is in burst frequency, the average burst size is constant, and 
we require that the constitutive protein have the same average 
burst size, adjusting its burst frequency to achieve the required mean. 
On the other hand, if feedback is in burst size, then
the burst frequency is constant, and we require that the constitutive
protein has the same burst frequency. This difference leads to
different constitutive CV$^2$'s and and hence different definitions of
${\rm CV}^2_{\rm rel}$ in the two cases: compare \eqref{cv2_rel} and 
\eqref{cv2_rel_2}.

\begin{figure}
\begin{center}
 \includegraphics[width=10cm]{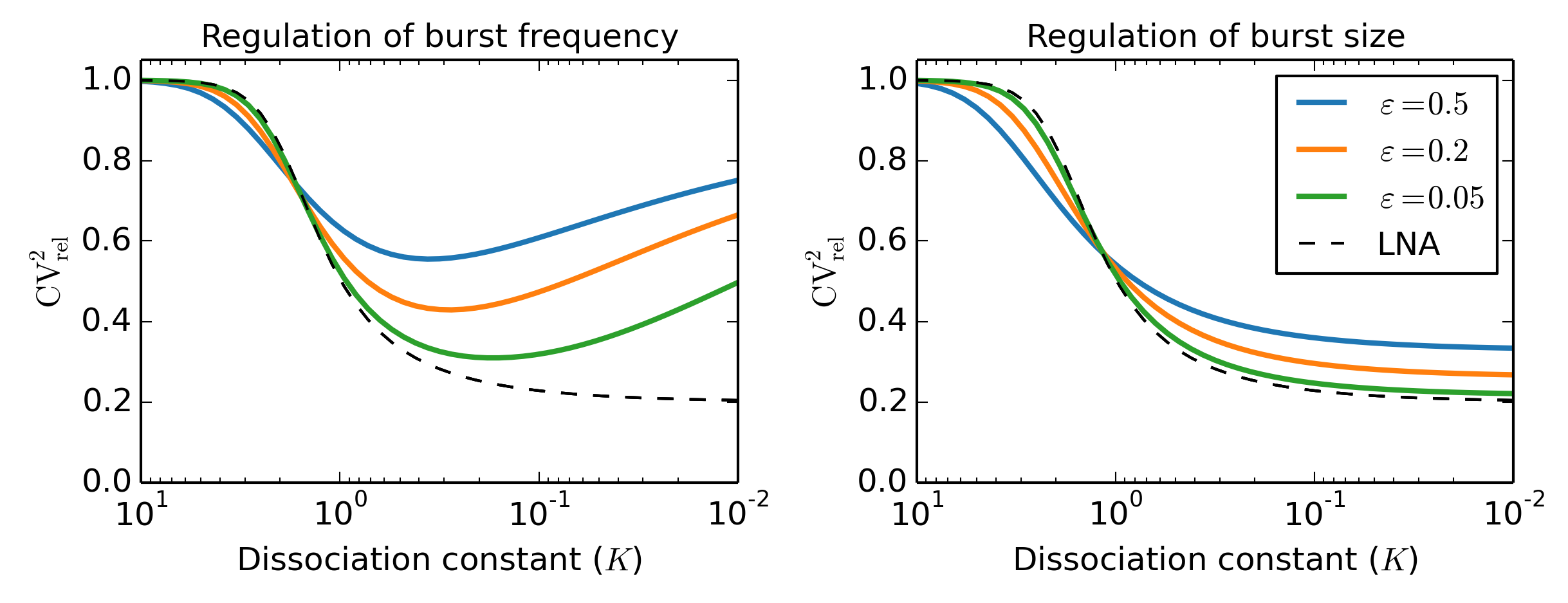}
\end{center}
\caption{Relative squared coefficient of variation (${\rm CV}^2_{\rm rel}$), i.e. the ratio of the 
regulated protein CV$^2$ relative to that of a constitutive protein, for feedback in burst frequency and size.}
\label{fig:cv2_rel}
\end{figure}

\paragraph{The two feedback types exhibit the same relative noise attenuation
in the small noise regime.} In the regime of small but frequent
bursts ($\ve\ll1$), linear noise approximation yields an explicit
expression for ${\rm CV}^2_{\rm rel}$. Interestingly, the 
same result is obtained whether feedback is in burst size or frequency, 
cf.~\eqref{cv2_lna} and~\eqref{cv2_lna_2}. The asymptotic
${\rm CV}^2_{\rm rel}$ decreases with increasing feedback
strength, converging to $1/(H+1)$ as the dissociation constant 
$K$ tends to zero (Figure~\ref{fig:cv2_rel}, both panels, dashed line).

\paragraph{Feedback in burst size outperforms feedback in burst frequency in
reducing relative noise outside of the small noise regime.} Unlike
in the small-noise regime, at moderate noise levels feedback type
influences ${\rm CV}^2_{\rm rel}$. While feedback of either type
brings about a decrease of ${\rm CV}^2_{\rm rel}$, feedback
in burst frequency is most efficient at intermediate strengths,
after which ${\rm CV}^2_{\rm rel}$ begins to increase again
(Figure~\ref{fig:cv2_rel}, left panel, solid coloured lines);
on the other hand, the response of ${\rm CV}^2_{\rm rel}$ to 
strengthening feedback in burst size is monotone, albeit less sigmoidal 
than in the LNA regime (Figure~\ref{fig:cv2_rel}, right panel, solid
coloured lines).

\paragraph{Noise optimisation for feedback in burst frequency.}
There is a positive value of the dissociation constant $K$ which minimises 
${\rm CV}^2_{\rm rel}$ for feedback
in burst frequency (Fig~\ref{fig:cv2_rel}, left). If feedback
is very weak ($K\gg1$), protein fluctuations are largely free of 
repression, and protein noise is close to that of an unregulated protein (i.e. ${\rm CV}^2_{\rm rel}\approx1$). 
If feedback is very strong ($K\ll\ve$), then protein time traces consist of 
separated, regularly spaced, bursts: each burst produces
amount of protein that is typically much larger than $K$, so that
immediately after the burst the propensity for another burst drops 
dramatically; only after the present amount of protein degrades 
another burst may occur. In contrast to an unregulated protein produced 
with the same average burst frequency, 
stringently self-repressed protein maintains regular time-spacing between 
individual bursts. The effect of regular bursting on protein noise becomes 
less important as feedback strength increases further, since stronger feedback 
implies a lower average burst frequency and hence a lesser chance of two bursts 
occurring at similar times by chance in the absence of regulation. Thus, in the 
strong-feedback limit of $K\rightarrow0$, noise of a strongly self-repressed protein 
is equal to that of an unregulated protein. In the intermediate range of dissociation 
constants $1\precsim K \precsim \ve$, negative feedback efficiently reduces 
protein fluctuations, and a regulated protein is less noisy than its 
unregulated counterpart.

\section{Discussion}
\label{sec:discuss}

In this paper we aimed to contribute towards the theoretical 
understanding of the effects of negative feedback on stochastic 
gene expression. Previous studies used small-noise approximations to obtain 
tractable expressions for protein noise characteristics as
functions of biochemical parameters~\citep{swain2004efficient,singh2011genetic, 
singh2011negative,singh2009reducing}. Others obtained exact,
but perhaps harder to interpret, results, which are valid
even at low molecule copy numbers or large-deviation 
regimes~\citep{hornos2005self,zeiser2009hybrid, fournier2007steady,grima2012steady}. We decided to combine the two approaches, comparing
the exact numerical predictions with asymptotic approximations
to obtain a complete characterisation for a minimalistic
model for the a protein produced in bursts subject to negative 
feedback.

Our results reinforce previously made observations that
downstream feedback (here feedback in burst size)
can better perform than upstream feedback (here regulation
of burst frequency) in reducing protein variability~\citep{swain2004efficient,singh2011genetic, 
singh2011negative,singh2009reducing,bandiera2016experimental}. For 
a protein which regulates its burst frequency, increasing 
feedback strength tends to increase the coefficient of variation, 
after an initial decrease. On the other hand, strengthening 
feedback in burst size leads to a monotone decrease in noise.

If instead of focusing on absolute coefficients of variation
we measure how does the feedback improve in reducing noise
on the performance of an equivalent constitutively expressed
protein, we obtain a subtler difference between the two
types of feedback, which is indeed indistinguishable in the
small-noise regime. However, outside of this regime,
even this subtler comparison shows a preference for 
regulation in burst size, especially in stringent feedback 
regimes. Hence, our approach suggests a possible role
of large deviations in distinguishing between the two
regulation mechanisms.

Our paper confirms the useful role asymptotic analysis can 
play in understanding the minutiae of stochastic gene 
expression~\citep{kuehn2015multiple,popovic2016geometric,bokes2012multiscale,newby2015bistable}. Asymptotics complements numerics, one working
well in parameter regimes where the other fails and vice
versa. More than one asymptotic regime may be needed to 
be considered in a given modelling context; our example
required two: small noise regime and strong feedback
regime. Finding the asymptotics in this two regimes
and filling the middle ground with numerical results
yielded a satisfactory understanding of the model behaviour
across the parameter space.

\renewcommand{\theequation}{A\arabic{equation}}
%reset counter 
\setcounter{equation}{0}

\section*{Appendix A. Reduction to the deterministic limit}

In the main text we showed that, irrespective of whether the feedback acts
on the burst frequency or burst size, the steady-state mean $\langle x\rangle$
of the protein concentration tends in the small-noise limit $\ve\rightarrow0$ to the 
fixed point $x_{\rm s}$ of the ordinary differential equation~\eqref{deterministic}.
Here we provide a stronger result, showing that for both feedback types the master 
equation~\eqref{continuity}--\eqref{flux} reduces as $\ve\rightarrow0$ to Liouville's 
partial differential equation associated with the ordinary differential 
equation~\eqref{deterministic}. Thus, we show that both regulation strategies 
yield the same deterministic model in the limit of small noise.

For feedback in burst frequency~\eqref{normal} the probability flux~\eqref{flux} 
is given by
\begin{equation}
\label{j1}
 J = - x p(x,t) + \frac{1}{\ve}\int_0^x   \frac{{\rm e}^{-\frac{x-y}{\ve}} p(y,t) \der y}{1 + (y/K)^H}
\end{equation}
If $\ve$ is small, a dominant contribution to the integral in~\eqref{j1} comes from a neighbourhood
of the upper integration limit $y=x$. Following Watson's lemma~\citep{hinch1991perturbation}, we
extend the lower integration limit in~\eqref{j1} to $-\infty$ and use the
approximation
\begin{equation}
\label{taylor_truncated}
\frac{p(y,t)}{1 + (y/K)^H} \sim \frac{p(x,t)}{1 + (x/K)^H} \quad \text{for $y$ that is close to $x$},
\end{equation}
obtaining 
\begin{equation}
\label{j1_approx}
J \sim  - x p  + \frac{p}{1 + (x/K)^H}
\end{equation}
at the leading order; higher-order terms, which are not required for our present purposes, 
can be determined  by including in~\eqref{taylor_truncated} additional terms of the Taylor 
series expansion in $y$ around $x$. The right-hand side of~\eqref{j1_approx}, being the product
of the protein pdf and the right-hand side of the ODE~\eqref{deterministic}, gives the 
flux of probability induced by the drift of the deterministic model. Inserting~\eqref{j1_approx} into the probability 
conservation law~\eqref{continuity} yields a Liouville equation~\citep[p. 213]{schuss2009theory} --- a Chapman--Kolmogorov equation without diffusion or jumps --- 
whose solutions are time-dependent pdfs for a variable which evolves deterministically
according to~\eqref{deterministic}. Thus, the stochastic model with feedback in burst frequency given by the conservation 
law~\eqref{continuity} and~\eqref{j1} reduces as $\ve$ tends to zero to the deterministic model~\eqref{deterministic}.

If feedback acts on burst size~\eqref{abnormal}, the probability flux~\eqref{flux} simplifies to
\begin{equation}
 \label{j2}
 J = - x p(x,t) + \frac{1}{\ve}\int_0^x {\rm e}^{-\frac{1}{\ve}\int_y^x 1 + (z/K)^H\der z} p(y,t)\der y.
\end{equation}
Again, a neighbourhood of the upper limit $y=x$ of integration dominates in its contribution to the integral in~\eqref{j2}. 
Therefore, we extend the lower integration limit to $-\infty$ without incurring appreciable error; we 
also use the approximations
\begin{equation}
\label{taylor2}
 \int_y^x 1 + (z/K)^H \der z \sim (1 + (x/K)^H)(x-y),\quad p(y,t) \sim p(x,t),
\end{equation}
which are valid for $y$ that is close to $x$. Inserting~\eqref{taylor2} into~\eqref{j2} and 
integrating the simple exponential, we obtain the same leading-order approximation~\eqref{j1_approx} 
for the probability flux~\eqref{j2} as we previously did for the flux~\eqref{j1}. Thus, whether the bursting 
stochastic model operates a feedback in burst frequency~\eqref{j1} or in burst size~\eqref{j2}, it reduces to 
the same deterministic model~\eqref{deterministic} in the small-noise limit of $\ve\rightarrow0$.

\renewcommand{\theequation}{B\arabic{equation}}
%reset counter 
\setcounter{equation}{0}

\section*{Appendix B. Strong feedback asymptotics (burst size)}

Inserting the second Lyapunov function~\eqref{phi2} into the WKB form~\eqref{wkb},
we obtain
\begin{equation}
 p(x) = C x^{\frac{1}{\ve}-1} {\rm e}^{-\frac{1}{\ve}\left(\frac{x^{H+1}}{(H+1)K^H} + x\right)}
\end{equation}
for the protein pdf in the case of feedback in burst size.

Similarly as in the Main Text, we express the protein moments as
\begin{equation}
\label{mudrost_sediva}
 \langle x^n \rangle = \frac{B_n}{B_0},
\end{equation}
where instead of~\eqref{bn} we have
\begin{equation}
\label{bn2}
 B_n = \int_0^\infty x^{\frac{1}{\ve} - 1 + n} {\rm e}^{-\frac{1}{\ve}\left(\frac{x^{H+1}}{(H+1)K^H} + x\right)} \der x.
\end{equation}
Again, $B_0^{-1}=C$ is the normalisation constant. Substituting $x=Ky$ in the integral~\eqref{bn2} yields
\begin{equation}
\label{bn_and_an2}
 B_n = K^{\frac{1}{\ve}+n} A_n,
\end{equation}
where
\begin{equation}
\label{an2}
A_n = \int_0^\infty y^{\frac{1}{\ve} - 1 + n} {\rm e}^{-\lambda\left( \frac{y^{H+1}}{H+1} + y \right)} \der y,
\end{equation}
in which $\lambda=K/\ve$ is an auxiliary parameter.

In the case of strong feedback, we have $\lambda\ll1$, which implies $y\gg1$, and therefore the term
$y^{H+1}/(H+1)$ dominates the term $y$ in the exponential of~\eqref{an2}, so that
\begin{equation}
\label{approximate}
A_n \sim \int_0^\infty y^{\frac{1}{\ve} - 1 + n} {\rm e}^{-\frac{\lambda y^{H+1}}{H+1}} \der y
      = \frac{1}{\lambda}\left(\frac{H+1}{\lambda}\right)^{\frac{\ve^{-1} + n}{1+H} - 1} \Gamma\left(\frac{\ve^{-1} + n}{1+H}\right),
\end{equation}
where $\Gamma(z)$ is the gamma function~\cite{abramowitz1972hmf}. Unlike for feedback in burst frequency, 
here is no need to treat $A_0$ differently from $A_1$ or $A_2$.

For the protein mean we have
\begin{equation}
 \langle x \rangle = \frac{B_1}{B_0} = \frac{K A_1}{A_0} \sim D_\ve K^{\frac{H}{H+1}},
\end{equation}
where the prefactor $D_\ve$ is given by
\begin{equation}
\label{prefactor}
 D_\ve = (\ve(H+1))^{\frac{1}{H+1}}\frac{\Gamma\left(\frac{\ve^{-1} + 1}{H + 1}\right)}{\Gamma\left(\frac{\ve^{-1}}{H + 1}\right)}.
\end{equation}
Unlike for feedback in burst frequency, which yielded a slow logarithmic decrease of the mean with
decreasing dissociation constant $K$, here we obtain a faster power-law 
decrease, which is consistent with the LNA prediction~\eqref{powerlaw}. Additionally, as $\ve$ tends 
to zero the prefactor $D_\ve$ converges to one, which is the prefactor of the LNA-predicted power law.
The asymptotics of $D_\ve$ as $\ve\rightarrow0$ follow from
\begin{equation}
\label{gamma_asymptotic}
 \frac{\Gamma(z + a)}{\Gamma(z)} \sim z^a,\quad z\gg1,
\end{equation}
in which we take $z=\ve^{-1}/(H+1)$ and $a=1/(H+1)$; see~\cite{abramowitz1972hmf} for this and other
properties of the gamma function. These results suggest that, unlike for feedback in burst frequency,
where the LNA approximation could only be used for intermediate ranges of $K$, here the LNA yields
a uniform (i.e. valid for all $K$) approximation.

For the protein CV$^2$ we have
\begin{equation}
 \label{cv2_asympt}
 {\rm CV}^2 = \frac{B_2 B_0}{B_1^2} - 1 = \frac{A_2 A_0}{A_1^2} - 1 \sim
 \frac{\Gamma\left(\frac{\ve^{-1}}{H+1}\right)\Gamma\left(\frac{\ve^{-1}+2}{H+1}\right)}{\Gamma^2\left(\frac{\ve^{-1}+1}{H+1}\right)} - 1,
\end{equation}
which, for a fixed $\ve$, is a constant independent of $K$. As $\ve$ tends to zero, we can again
use~\eqref{gamma_asymptotic} to show that the right-hand side of~\eqref{cv2_asympt}
is equal to $\ve/(H+1)$ at the leading order in $\ve$, which is the same value as that obtained by taking $K$ 
very small in the LNA prediction~\eqref{cv2_lna_2}. This suggests that the LNA approximation of the 
coefficient of variation, like that of the mean, can be used uniformly for all $K$.

\renewcommand{\theequation}{C\arabic{equation}}
\renewcommand{\thefigure}{C\arabic{figure}}
%reset counter 
\setcounter{equation}{0}
\setcounter{figure}{0}

\section*{Appendix C. Noncooperative feedback}

Here we present variants of the figures from the Results in the Main Text
obtained by taking $H=1$ (noncooperative feedback) instead of $H=4$.
We shall not repeat the points made in the Main Text, focusing instead
on the main differences that occur in the absence of cooperativity.

\begin{figure}
\begin{center}
\includegraphics[width=10cm]{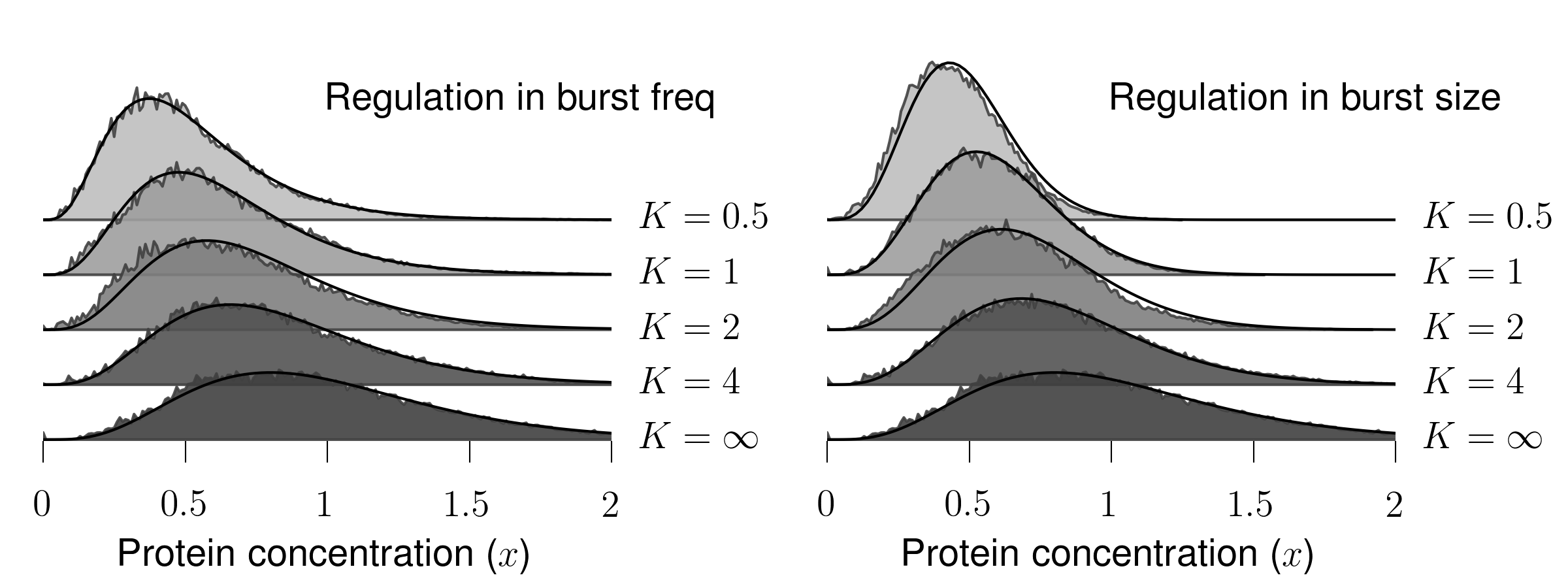}
\end{center}
\caption{Protein distributions for varied feedback strength. $\ve=0.2$.}
\label{fig:distributionsh1}
\end{figure}

\begin{figure}
\begin{center}
 \includegraphics[width=10cm]{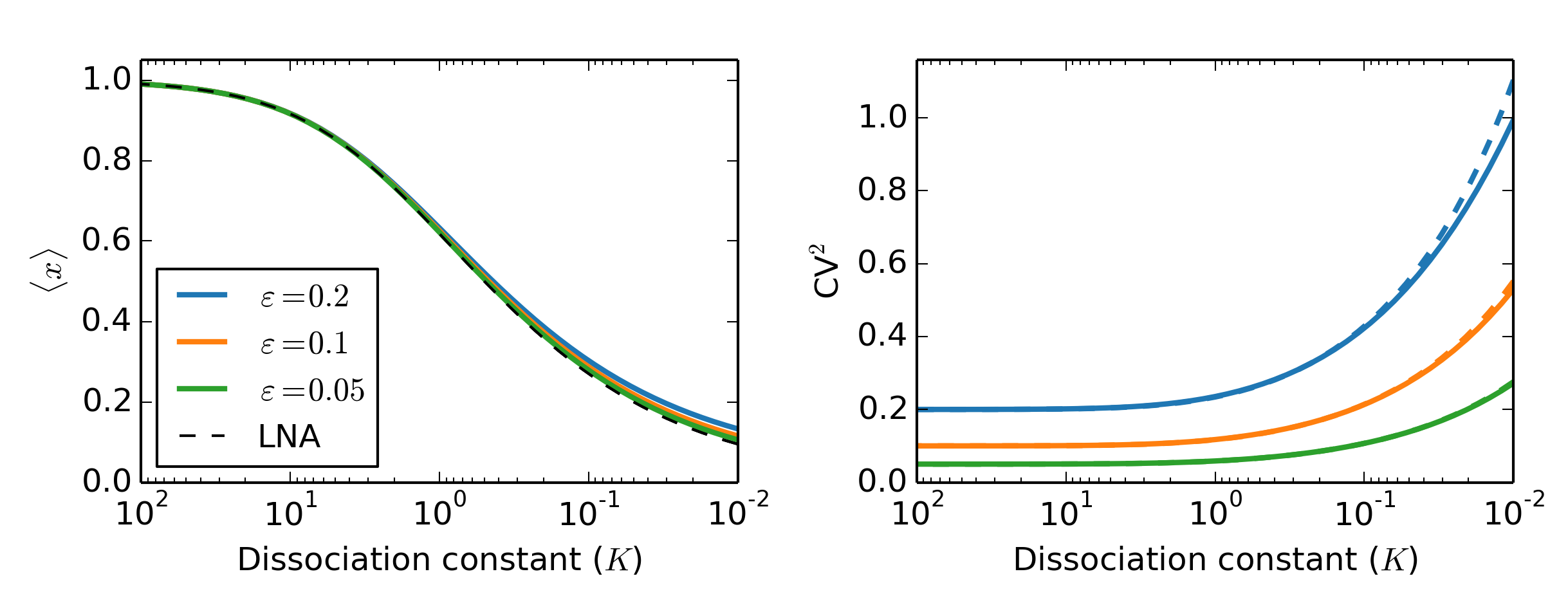}
\end{center}
\caption{Protein mean and CV$^2$ in response to strengthening feedback in burst frequency.}
\label{fig:cv2_bfreq_regh1}
\end{figure}

\begin{figure}
\begin{center}
\includegraphics[width=10cm]{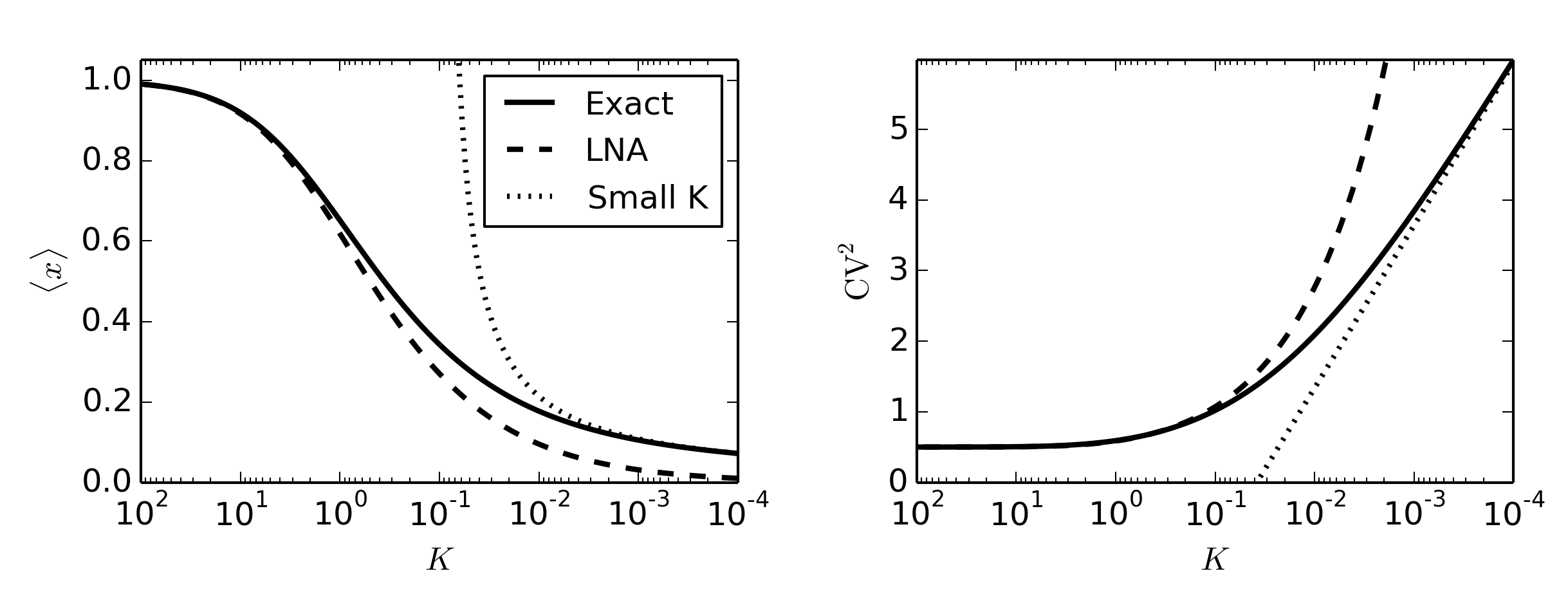}
\end{center}
\caption{Impact of strengthening feedback in burst frequency on mean and CV$^2$ of a noisy protein ($\ve=0.5$). 
Comparison of exact results (full line) with LNA (dashed line) and small $K$ (dotted line) asymptotics.}
\label{fig:asymptoticsh1}
\end{figure}

\begin{figure}
\begin{center}
 \includegraphics[width=10cm]{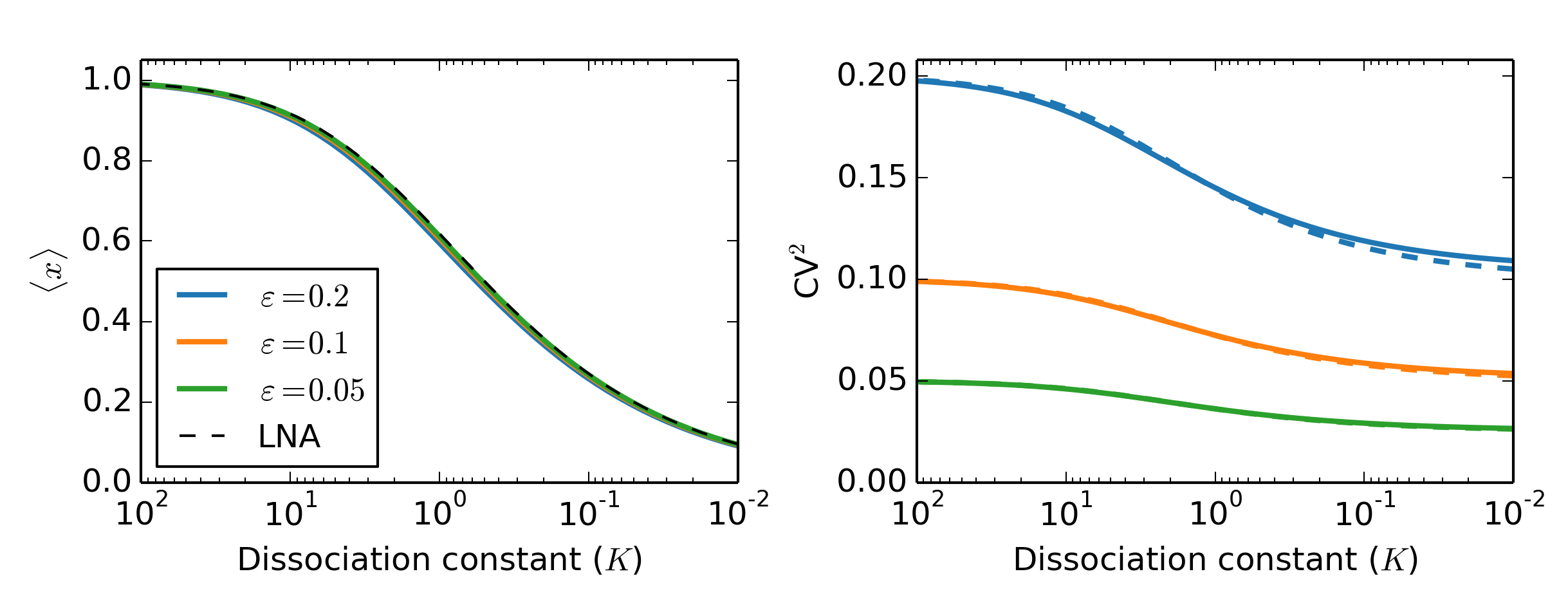}
\end{center}
\caption{Protein mean and CV$^2$ in response to strengthening feedback in burst size.}
\label{fig:cv2_bsize_regh1}
\end{figure}

\begin{figure}
\begin{center}
 \includegraphics[width=10cm]{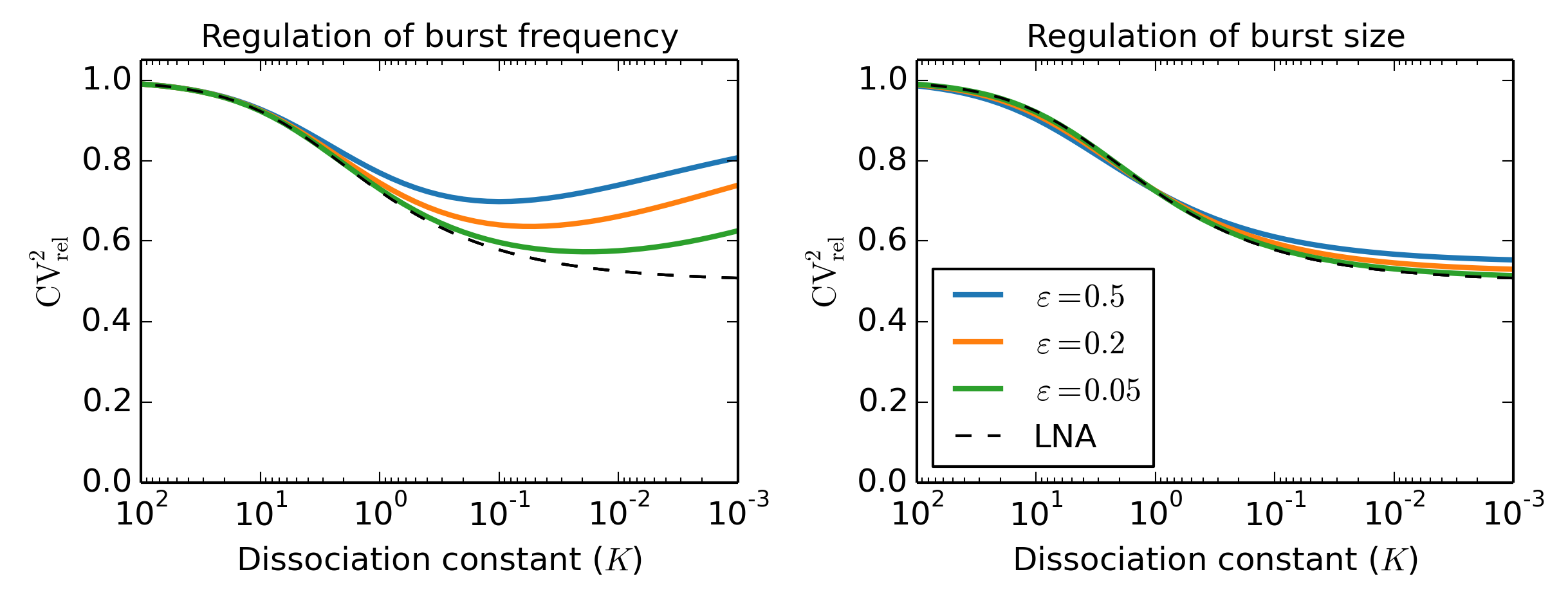}
\end{center}
\caption{Relative squared coefficient of variation (${\rm CV}^2_{\rm rel}$), i.e. the ratio of the 
regulated protein CV$^2$ relative to that of a constitutive protein for feedback in burst frequency and size.}
\label{fig:cv2_relh1}
\end{figure}

{\bf CV$^2$ monotonically increases with strengthening noncooperative 
feedback in burst frequency.} In contrast with the cooperative case,
where a gradual increase in burst-frequency feedback strength led at first 
to a transient decrease in protein noise (Fig~\ref{fig:cv2_bfreq_reg}), 
without cooperativity the CV$^2$ is strictly increasing (Fig~\ref{fig:cv2_bfreq_regh1}).

{\bf Protein mean and CV$^2$ are less sensitive to feedback strength.}
Wider ranges of dissociation constants are required to achieve 
similar changes in protein mean and noise as those reported
previously. In order to appreciate this, one needs to compare the scales on 
the horizontal axes of Figures~\ref{fig:cv2_bfreq_regh1}--\ref{fig:cv2_relh1}
with those of their counterparts in the Main Text.

{\bf Noncooperative performs worse than the cooperative in reducing noise.}
Noncooperative feedback, even if acting through burst size, leads at best to
a $50\%$ reduction in CV$^2$ (Fig~\ref{fig:cv2_relh1}), which is inferior to 
a $80\%$ reduction achievable in the cooperative case with $H=4$ 
(Fig~\ref{fig:cv2_rel}).

{\bf The main conclusion of the Main Text holds also in noncooperative case.}
The regulation in burst size performs better in reducing noise, especially
for noisy proteins subject to strong self-repression (Fig~\ref{fig:cv2_relh1}).

\renewcommand{\theequation}{D\arabic{equation}}
\renewcommand{\thetable}{D\arabic{table}}
%reset counter 
\setcounter{equation}{0}
\setcounter{table}{0}

\section*{Appendix D. Discrete simulations}

\begin{table}[b]
\begin{center}
\begin{tabular}{lcr}
Reaction name & Copy number change & Stochastic rate \\
\hline\\[-0.5em]
Activation & $A \rightarrow A + 1$ & $(1-A)\tilde{k}_{\rm on}(P)$ \\
Inactivation & $A \rightarrow A - 1$ & $A\tilde{k}_{\rm off}(P)$ \\
Protein production & $P \rightarrow P + 1$ & $A\tilde{k}_{\rm p}(P)$ \\
Protein decay & $P \rightarrow P - 1$ & $\tilde{k}_{\rm d}(P)$ \\
\end{tabular}
\end{center}
\caption{Reactions, their stoichiometries, and rates for the 
discrete stochastic model.}
\label{table:reactions}
\end{table}

The discrete model is a chemical system of two species~\cite{wilkinson2006sms}, 
$A$ and $P$, whereby $A\in\{0,1\}$ is an indicator variable describing whether
the gene is active ($A=1$) or inactive ($A=0$) and $P$ gives the
number of protein.

The two species are subject to four reactions, gene activation,
gene inactivation, protein production, and protein decay. Each
reaction is characterised by the change in copy numbers that a
single occurrence of the reaction induces and by the stochastic
rate with which the reaction occurs (Table~\ref{table:reactions}).

The dependence of the rates of activation $\tilde{k}_{\rm on}(P)$,
inactivation $\tilde{k}_{\rm off}(P)$, protein production $\tilde{k}_{\rm p}(P)$
and protein decay $\tilde{k}_{\rm d}(P)$ is as yet undefined in Table~\ref{table:reactions}, 
but is specified below for feedbacks in burst frequency and burst size. 
We use tildes
to distinguish the microscopic rates (expressed in terms of individual 
molecules) from the macroscopic ones (expressed in terms of concentrations)
which were used throughout the Main Text.

For feedback in burst frequency, we choose
\begin{equation}
 \label{normal_discrete}
 \tilde{k}_{\rm on}(P)= \frac{\ve^{-1}}{1 + (P/K\Omega)^H},\quad
  \tilde{k}_{\rm p}(P) = \frac{\ve\Omega}{\delta},\quad
 \tilde{k}_{\rm off}(P) = \frac{1}{\delta},\quad
 \tilde{k}_{\rm d}(P) = P,
\end{equation}
while for feedback in burst size, we use
\begin{equation}
 \label{abnormal_discrete}
 \tilde{k}_{\rm on}(P)= \ve^{-1},\quad
  \tilde{k}_{\rm p}(P) = \frac{\ve\Omega}{\delta(1 + (P/K\Omega)^H)},\quad
 \tilde{k}_{\rm off}(P) = \frac{1}{\delta},\quad
 \tilde{k}_{\rm d}(P) = P.
\end{equation}
In addition to the noise parameter $\ve$, dimensionless dissociation constant $K$ and
the cooperativity coefficient $H$, which have been introduced in the Main Text, 
cf. Eq.~\eqref{normal} and~\eqref{abnormal}, we have in~\eqref{normal_discrete} 
and~\eqref{abnormal_discrete} two new parameters: $\delta$ and $\Omega$. The parameter 
$\delta$ compares the time scale of gene activity to that of protein turnover, and $\Omega$ 
is the system size parameter: the number of proteins corresponding to the unit of 
concentration.

Provided that $\delta\ll1$ and $\Omega\gg1$, the protein concentration defined
as $x=P/\Omega$ can be compared to the predictions of the continuous bursting
model~\eqref{ss}. For mathematical analysis of the bursting asymptotics ($\delta\ll1$) 
as well as system-size asymptotics ($\Omega\gg1$), we refer the reader 
to~\cite{bokes2012multiscale}.

In Figure~\ref{fig:distributions}, we used $\ve=0.2$, $H=4$, a range
of values of $K$ (detailed within the figure panels), $\delta=0.01$ and $\Omega=100$.
Each distribution was estimated from a single large run ($10^5$ iterations) of 
Gillespie's direct method~\cite{gillespie1977ess} implemented in the StochPy 
stochastic modelling software package~\citep{maarleveld2013stochpy}.

\section*{Acknowledgements}

PB is supported by the Slovak Research and Development Agency grant APVV-14-0378 and 
also by the VEGA grant 1/0319/15. AS is supported by the National Science Foundation grant DMS-1312926. 
The authors thank Daniel \v{S}ev\v{c}ovi\v{c} and Branislav Novotn\'{y} for discussion on some 
of the ideas contained herein.


\bibliographystyle{unsrtnat}     
\bibliography{autonew,refs_selection}

\end{document}